\documentclass[epj,a4paper]{svjour}

\usepackage{amsmath}
\usepackage{latexsym}
\usepackage{graphicx}

\newcommand{\bra}[1]{\langle#1|}
\newcommand{\ket}[1]{|#1\rangle}

\begin{document}

\title{Charge profile of surface doped C$_{60}$}

\author{Samuel Wehrli\inst{1}\thanks{\email{swehrli@itp.phys.ethz.ch}} \and 
        Didier Poilblanc\inst{1,2} \and T.\ M.\ Rice\inst{1}}

\institute{Theoretische Physik, Eidgen\"{o}ssische Technische Hochschule,
           CH-8093 Z\"{u}rich,
           Switzerland \and
           Laboratoire de Physique Quantique \& UMR--CNRS 5626, Universit\'e
           Paul Sabatier, F-31062 Toulouse, France}

\date{\today}

\abstract{ 
  We study the charge profile of a C$_{60}$-FET 
  (field effect transistor) as used
  in the experiments of Sch\"on, Kloc and Batlogg. Using a tight-binding model,  we calculate the charge profile treating
  the Coulomb interaction in
  a mean-field approximation. At low doping, the 
  charge profile behaves similarly to the
  case of a continuous
  space-charge 
  layer and becomes confined to
  a single interface layer for doping
  higher than $\sim 0.3$ electron (or hole) per C$_{60}$ molecule. 
  The morahedral disorder of the C$_{60}$ molecules smoothens the structure in
  the density of states.
}

\maketitle


Surface doping of high quality organic crystals has recently been achieved by
Sch\"on, Kloc and Batlogg~\cite{art:schoen0,art:schoen0b,art:schoen0c}. This
breakthrough has led to novel devices, new superconductors and measurements of
transport properties over unprecedented wide ranges of carrier concentration
and temperature. A tunable field effect transistor (FET) is used to inject
carriers near to the surface of the organic crystal. In this paper we report
on calculations of the resulting charge profile. We chose to investigate the
case of a C$_{60}$-crystal which has a special interest in view of the recent
observations of superconductivity with a high $T_c$ ($\sim 52$~K) in hole doped
samples. 

\begin{figure}
\begin{center}
(a)\hspace{0.5cm}
\begin{minipage}[c]{0.35\textwidth}
  \begin{center}
    \includegraphics[width=\textwidth]{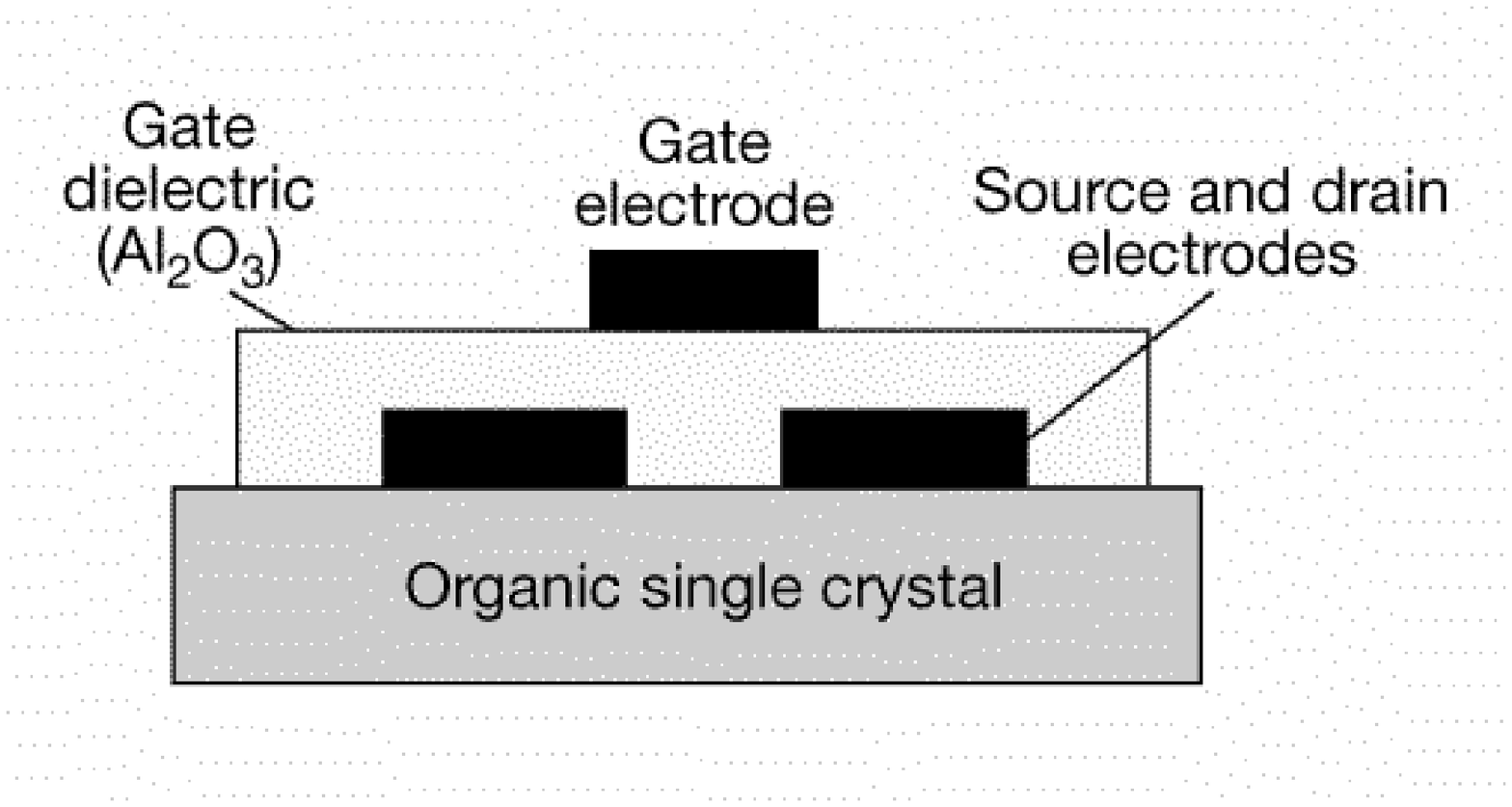}
  \end{center}
\end{minipage}\vspace{0.3cm}\\
(b)\hspace{0.5cm}
\begin{minipage}[c]{0.35\textwidth}
  \begin{center}
    \includegraphics[width=\textwidth]{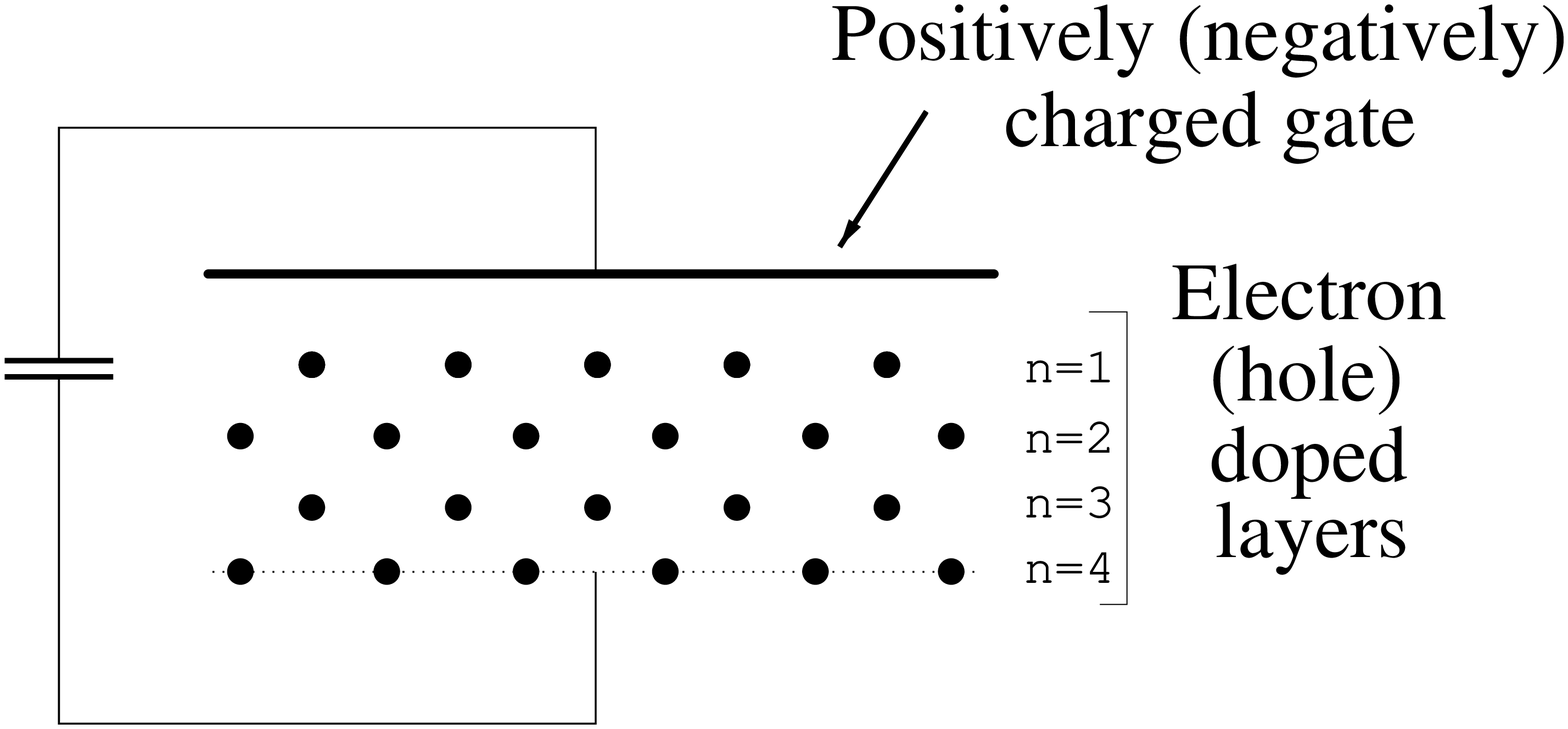}
  \end{center}
\end{minipage}\vspace{0.3cm}\\
\caption{
(a) Schematic picture of the FET as used in the 
experiments of Sch\"on, Kloc and Batlogg. Picture taken
from~\cite{art:schoen0}. 
(b) Model of the electronic system in the FET.
}\label{fig:FET_Device}
\end{center}
\end{figure}

The FET used in the work of
Sch\"on et al. is illustrated in Fig.~\ref{fig:FET_Device}(a). In the
present work, we consider a C$_{60}$ crystal (fcc lattice) 
with a [001] plane parallel
 to the gate.
Undoped  C$_{60}$
is a semiconductor with a 2~eV gap. However,
when an electric potential is applied between the gate electrode and the
source/drain electrodes, either electrons or holes accumulate on the interface
between the C$_{60}$ crystal and the gate dielectric. This
leads to a doping of the C$_{60}$ interface planes which allows
current to flow parallel to the interface, between the source and the drain. The resistance of this
channel can be measured as a function of temperature and doping. 
The FET can be represented as a planar capacitance with equal and opposite
charges on the metal gate and the C$_{60}$ planes near to the interface as
shown in Fig.~\ref{fig:FET_Device}(b). We
will calculate the charge profile as a function of the total charge induced at
the interface (or equivalently on the metal gate) rather then in terms of the
potential applied to the gate. We limit ourselves to the case of an ideal
planar interface with no steps or imperfections.

We begin by introducing a tight-binding model to describe the electronic
structure of the   C$_{60}$ crystal. The effects of Coulomb interactions are
treated within a mean-field (or Hartree) approximation. The resulting density
profile gives rise to highly structured density of states (DOS). However we
show that the inclusion of morahedral (or orientational) disorder leads to a
broadening and suppression of this structure in the DOS.


The C$_{60}$ molecule, which has icosahedral symmetry, is almost spherical. 
Therefore, in the solid, it orders naturally in
the close packed fcc structure. In the following 
we will assume the (hypothetical)
unidirectional structure where all C$_{60}$
molecules have the same orientation~\cite{art:satpathy92}. The space group for
this structure is $Fm\bar 3m$ and its primitive cell contains one molecule.
The actual low-temperature structure with space group $Pa\bar 3$ 
is more complicated and has a unit cell with 4 molecules. However, in the
present work, we are interested in the 
overall charge filling of the layers, for which  details of the
band structure are unimportant.
The lattice constant of the cubic cell (containing two molecules) 
is $a=14$~\AA{} for pure C$_{60}$~\cite{art:liu91}. 
We introduce here the unit vectors
$\vec e_x$, $\vec e_y$ and $\vec e_z$ which
span the cubic cell.  If one considers only a [001] layer, then the C$_{60}$
molecules form a 2D square lattice. In this case
the primitive cell has a side length of
$b=\frac{a}{\sqrt 2}=10$\AA{} which is the distance between two neighboring
molecules. The 2D primitive cell is spanned by the unit vectors
$\vec e_1=\frac{1}{\sqrt2}(\vec e_x+\vec e_y)$ and
$\vec e_2=\frac{1}{\sqrt2}(-\vec e_x+\vec e_y)$.

\begin{figure}
\begin{center}
(a)\hspace{0.5cm}
\begin{minipage}[c]{0.3\textwidth}
  \includegraphics[width=\textwidth]{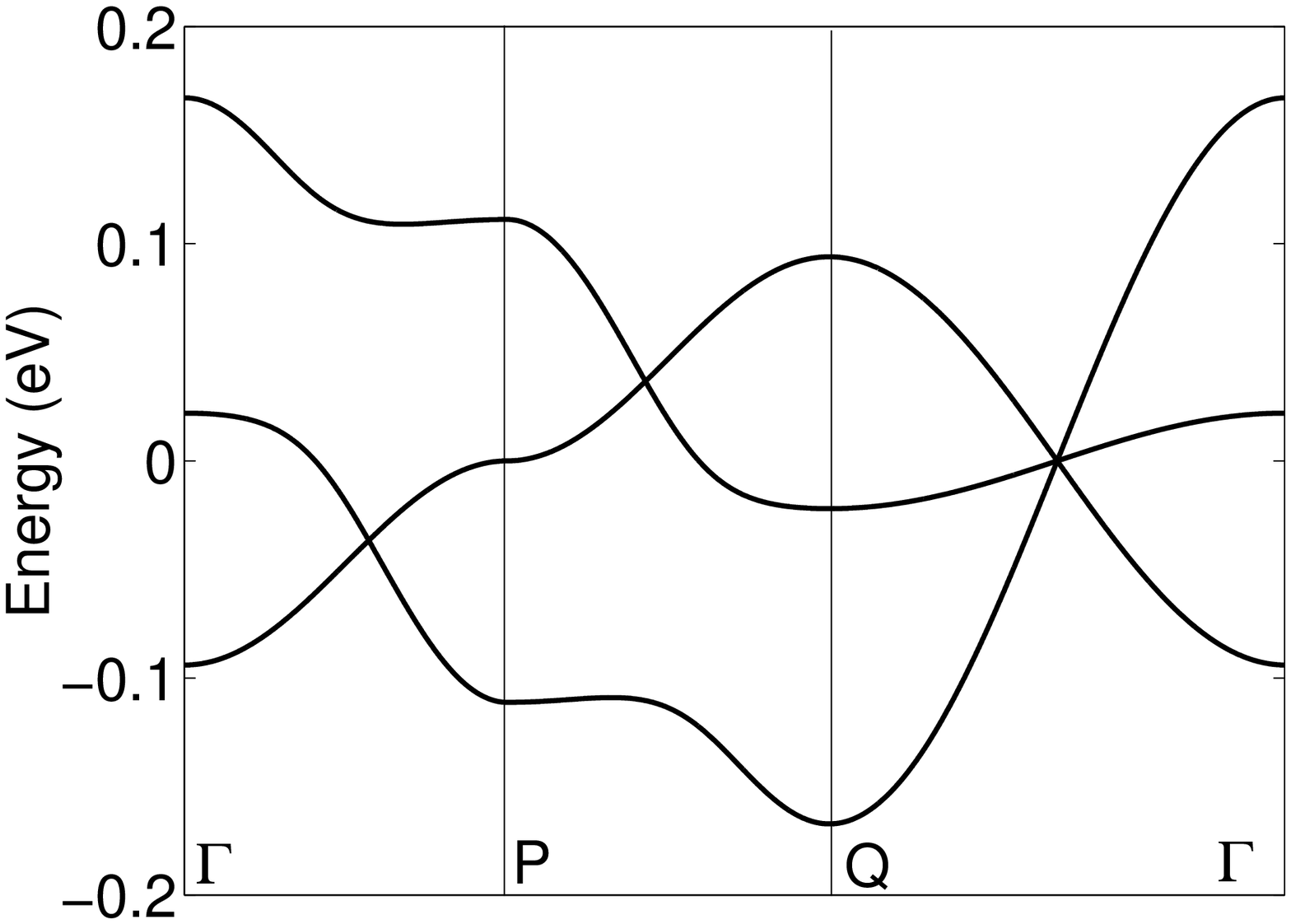}
\end{minipage}\\
(b)\hspace{0.5cm}
\begin{minipage}[c]{0.29\textwidth}
  \includegraphics[width=\textwidth]{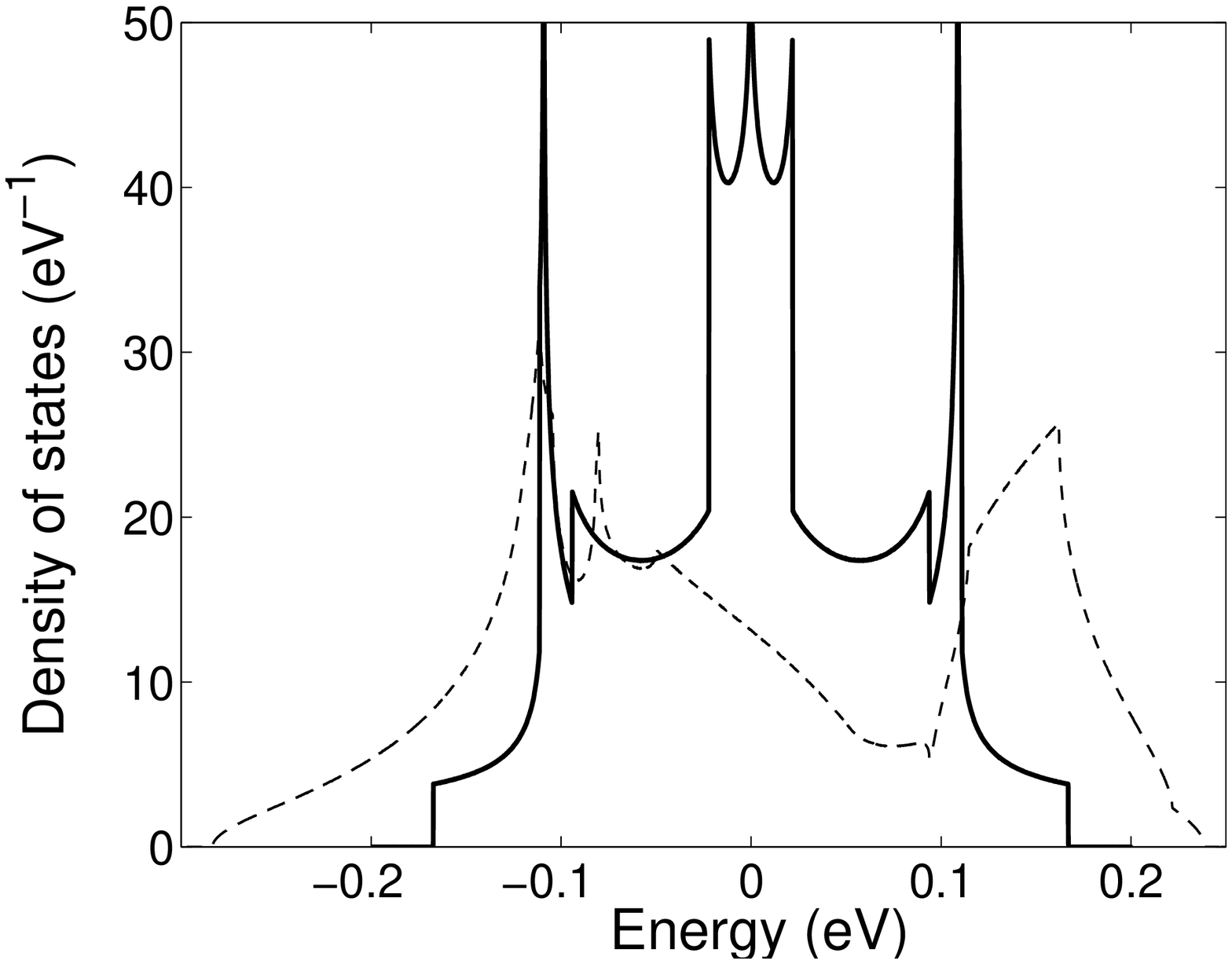}
\end{minipage}\vspace{0.5cm}\\
\caption{
(a)
C$_{60}$ conduction band of the 2D square lattice. The coordinates of the
  high-symmetry points in the Brillouin zone are $\Gamma(0,0)$,
  P$(\frac{\pi}{b},0)$,Q$(\frac{\pi}{b},\frac{\pi}{b})$.
(b)
DOS per C$_{60}$ and for both spins. \emph{Solid line:} 2
dimensions. \emph{Dashed line:} 3
dimensions. 
}\label{fig:CondBand}
\end{center}
\end{figure}

The kinetic energy is well described by a
tight-binding Hamiltonian with nearest neighbor (n.n.)
hopping~\cite{art:satpathy92,art:laouini95}: 
\begin{equation} \label{eq:TBHam}
  H_{\textrm{TB}} = \sum_{\langle nj\alpha,n'j'\alpha'\rangle}
    t_{\alpha'\alpha}(\vec{\delta})\;
      c_{n'j'\alpha'}^{\dagger}\; c_{nj\alpha}\; ,  
\end{equation}
where $t_{\alpha'\alpha}(\vec{\delta})$ are the hopping integrals
depending on the relative position 
$\vec{\delta}$ of
neighboring molecules. The label $n$ denotes the different layers
starting with the $n=1$ layer closest to the
gate. Within a layer $n$, the different sites are labelled by the index
$j$. Finally, the index $\alpha$ denotes the different orbitals of the
C$_{60}$ molecule.  The brackets $\langle\rangle$ indicate a summation
over n.n. only. Summation over spin degrees of freedom is  
implicitly assumed.
Hamiltonian~(\ref{eq:TBHam}) is invariant under translations parallel to
the $z=0$ plane and can partly be diagonalized in the 2D k-space:
\begin{equation} \label{eq:TBHam2}
  H_{\textrm{TB}} = \sum_{\vec{k}}
        \sum_{n\alpha,n'\alpha'}
    H_{n'\alpha',n\alpha}(\vec{k})\;
      c_{n'\alpha'}^{\dagger}(\vec{k})\; 
      c_{n\alpha}(\vec{k})\;, 
\end{equation}
where $|n-n'| \le 1$ and $\vec k$ is the two-dimensional wave-vector parallel
to the planes. In the case of the conduction band, the structure of the
hopping integrals 
$t_{\alpha'\alpha}(\vec{\delta})$ are 
given by the t$_{1u}$ symmetry of the
LUMO (Lowest Unoccupied Molecular Orbital) which is threefold
degenerate~\cite{art:laouini95}.  The corresponding wave-functions  
can be chosen such as to
transform as x, y and z under the icosahedral symmetry group which 
reduces the number of independent hopping
integrals to $4$.  
They  are given in Table~\ref{tab:HopInt}. The matrix elements 
$H_{n'\alpha',n\alpha}(\vec{k})$ in~(\ref{eq:TBHam2}) can be
calculated explicitly which yields for intraplanar processes ($n'=n$),
\begin{eqnarray} \label{eq:Ham2D}
  H_{nx,nx}(\vec{k}) &=  &2 t_{xx} [ \cos(k_1b) + \cos(k_2b)], \nonumber\\ 
  H_{ny,ny}(\vec{k}) &=   &2 t_{yy} [\cos(k_1b) + \cos(k_2b)],\nonumber\\ 
  H_{nx,ny}(\vec{k}) &= &2 t_{xy} [\cos(k_1b) - \cos(k_2b)],\\ 
  H_{nz,nz}(\vec{k}) &=  &2 t_{zz} [\cos(k_1b) + \cos(k_2b)],\nonumber
\end{eqnarray}
\noindent and for interplanar processes ($n'=n+1$),
\begin{eqnarray} \label{eq:Ham2DCpl}
 H_{n\!+\!1x,nx}(\vec{k})&= 
     &2t_{yy}\cos[(k_1\!\!-\!\!k_2)b/2]\!\! +\!\! 
      2t_{zz}\cos[(k_1\!\!+\!\!k_2)b/2], \nonumber\\ 
 H_{n\!+\!1y,ny}(\vec{k})&= 
     &2t_{zz}\cos[(k_1\!\!-\!\!k_2)b/2]\!\! +\!\! 
      2t_{xx}\cos[(k_1\!\!+\!\!k_2)b/2], \nonumber\\ 
 H_{n\!+\!1z,nz}(\vec{k}) &= 
     &2t_{xx}\cos[(k_1\!\!-\!\!k_2)b/2] \!\!+\!\! 
      2t_{yy}\cos[(k_1\!\!+\!\!k_2)b/2],  \nonumber\\
 H_{n\!+\!1x,nz}(\vec{k}) &=   &i\,2t_{xy}\sin[(k_1-k_2)b/2], \\ 
 H_{n\!+\!1y,nz}(\vec{k}) &=  &i\,2t_{xy}\sin[(k_1+k_2)b/2]. \nonumber 
\end{eqnarray}
The band structure and DOS of a single layer (interplanar processes are turned
off) are shown in
Fig.~\ref{fig:CondBand}. As a comparison, the 3D DOS is shown as well. The logarithmic van-Hove singularities are
clearly apparent in the DOS. Moreover, one observes a symmetric DOS. This
is explained by the substitution
$\vec k\to \vec k+(\frac{\pi}{b},\frac{\pi}{b})$ 
which leads to $H\to -H$ for the one-layer
Hamiltonian~(\ref{eq:Ham2D}). It implies also that all bands at
$(\frac{\pi}{2b},\frac{\pi}{2b})$ cross at zero energy. 
Coupling to neighboring layers breaks this symmetry.
The structure of the hopping integrals for the valence band are somewhat more
complicated  due to the fivefold degeneracy of the HOMO (Highest Occupied
Molecular Orbital) with h$_u$ symmetry. They are described in detail in
Ref.~\cite{art:laouini95} and similar matrix elements as in~(\ref{eq:Ham2D})
and~(\ref{eq:Ham2DCpl}) were calculated with parameters 
taken from~\cite{art:laouini95}.
\begin{table} 
\begin{center}
\begin{tabular}{c|ccc}
&  $\ket{x}$ & $\ket{y}$ &  $\ket{z}$ \\
\hline
$\bra{x}$ & $t_{xx}=5.5$  & $\pm t_{xy}=-27.8$  & $0$ \\
$\bra{y}$ &  $\pm t_{xy}=-27.8$  & $t_{yy}=41.8$         & $0$ \\
$\bra{z}$ & $0$              & $0$           & $t_{zz}=-23.5$  \\
\end{tabular}
\caption{
Hopping integrals (in meV) for hopping in the $\vec{\delta}=(110)$
 direction of the 
$[\vec e_x\vec e_y\vec e_z]$ coordinate
system. The sign of $t_{xy}$ changes for hopping in the
$(1\bar{1}0)$ direction. Hopping integrals in other directions follow from
rotations around the threefold (111) axes.
 Numerical values were taken from~\cite{art:satpathy92}
 and adjusted by a factor 1.11 for a  lattice
constant
$a=14.0$~\AA{} of pure C$_{60}$.  }\label{tab:HopInt}
\end{center}
\end{table}

The Coulomb interaction can be included 
by adding the following term to the tight-binding
Hamiltonian~(\ref{eq:TBHam}),
\begin{eqnarray}  \label{eq:HCoul}
   H_{\textrm{Coul}} &=& 
     \sum_{nj\alpha} U_{\textrm{Gate}}(n) n_{nj\alpha}  \\
     & &+ \frac{1}{2}\sum_{nj\alpha, n'j'\alpha'}
     V_{ee}(nj,n'j')n_{nj\alpha}n_{n'j'\alpha'}, \nonumber
\end{eqnarray}
where $n_{nj\alpha}$ is the number operator.
The first term is the potential due to the positive charge 
on the gate. The second term
is the electron-electron 
Coulomb repulsion. Here we follow Antropov et
al.~\cite{art:antropov92} by using a screened
Coulomb interaction for electrons on different sites and an on-site interaction
$U_0$ for electrons on the same site, 
\begin{equation}  \label{eq:Vee}
   V_{ee}(nj,n'j') = 
     \left\{ \begin{array}{cl}
                U_0 & \quad \textrm{if}\quad \vec R_{nj} =  \vec R_{n'j'} \\ 
                \frac{e^2}{\varepsilon 
                    \left|\vec R_{nj}- \vec R_{n'j'} \right|}
                   &\quad \textrm{if}\quad \vec R_{nj} \ne  \vec R_{n'j'} \\ 
              \end{array} \right. \; ,
\end{equation}  
with $\varepsilon$ being the dielectric constant of C$_{60}$ and $\vec R_{nj}$
denoting  the positions of the sites. Note that this interaction does not distinguish between
different orbitals. The dielectric constant
$\varepsilon$  and the
on-site interaction $U_0$ are easily accessible in the literature.
We use the experimental estimation of $\varepsilon=4.4$ by Hebard et 
al.~\cite{art:hebard91b} and the theoretical value of $U_0=1$~eV by
Antropov et al.~\cite{art:antropov92}. Somewhat different values of
$\varepsilon=3.66$ and $U_0=1.27$~eV are proposed from theoretical calculations
by Pederson et al.~\cite{art:pederson92}.


The full Hamiltonian including the Coulomb term is a
complicated many-body problem. We solve it within the
mean-field approximation. From symmetry, solutions should be
 invariant under translations parallel to the planes. The mean-field charge
 distribution and the resulting potential are then
independent of the in-plane site position $j$ and the
wave-functions can be decomposed as
\begin{equation} \label{eq:WF}
  \Psi_{\nu \vec k}(nj\alpha) = \frac{1}{\sqrt{N_\Box}}
    e^{i\vec R_{nj}^\Box \vec k}
      \psi_{\nu \vec k}(n\alpha),
\end{equation}
where $\nu$ is a band index and $\vec R_{nj}^\Box$ is the inplane component of
$\vec R_{nj}$. $N_\Box$ is the number of sites per layer.
For a fixed  $\vec k$, the perpendicular part $\psi_{\nu \vec
  k}(n\alpha)$ of the wave-function is the solution of a Schr\"odinger equation
with a partly discrete spectrum \{$E_{\nu \vec k}$\}.
In the mean-field approximation, every occupied electronic
${\nu \vec k}$ state contributes 
$\left|\psi_{\nu \vec k}(n\alpha)\right|^2/N_\Box$ electrons to
each orbital $\alpha$ in
layer $n$. The sum over all these contributions leads, at zero temperature, to
the electron distribution
\begin{equation} \label{eq:rho}
  \rho(n) = \frac{2}{N_\Box} \sum_{E_{\nu \vec k} < E_F}\sum_\alpha
     \left|\psi_{\nu \vec k}(n\alpha)\right|^2 ,
\end{equation} 
where $\rho(n)$ is the total number of electrons per site in layer $n$. The
factor  2 comes from the spin degrees of freedom. 
Within the mean-field approximation, the Coulomb part~(\ref{eq:HCoul}) can be
written as $H_{\textrm{Coul}}^{\textrm{MF}}=\sum_{nj\alpha}U_{\textrm{MF}}(n)
n_{nj\alpha}$ where
\begin{equation} \label{eq:UMF}
  U_{\textrm{MF}}(n)= U_{\textrm{Gate}}(n) +
    \sum_{n'j'} \, \rho(n') \, V_{ee}(nj,n'j') \, .
\end{equation} 
The sum in the
second term as well as the potential created by the gate are
divergent. However,  these divergences cancel when the overall system is
neutral. In order to make this more explicit, we add and subtract
a hypothetical potential created by uniformly 
charged layers whose total charge
equals the actual charge of the C$_{60}$ layers.
The mean-field potential can then be rewritten as a
sum of two terms
\begin{equation} \label{eq:UMF2}
  U_{\textrm{MF}}(n)= U_{\textrm{Cap}}(n)+U_{\textrm{Corr}}(n).
\end{equation} 
The first term,
$U_{\textrm{Cap}}$, is given by the superposition 
 of a series of planar
capacitance formed by the negative charge $-e\rho(n')$ on each layer $n'$ 
 combined
with the corresponding positive charge $+e\rho(n')$ on the gate. Choosing
$U_{\textrm{Cap}}(1)=0$ yields
\begin{equation} \label{eq:UCap2}
  U_{\textrm{Cap}}(n)=\frac{4\pi e^2}{\varepsilon \sqrt 2 b}
  \sum_{n'}\;\rho(n')\,[\min(n,n')-1],
\end{equation}
where it has to be distinguished whether the given layer n lies inside or outside the
capacitance with charges $\pm e\rho(n')$. 
Using $\varepsilon = 4.4$ and $b=10$~\AA{}
leads to a prefactor $4\pi e^2/\varepsilon \sqrt 2 b=2.9$~eV.
The second part of the mean-field potential, $U_{\textrm{Corr}}$, includes the
corrections arising from the discreteness of the charge distribution and is
given by
\begin{eqnarray} \label{eq:UCorr}
  U_{\textrm{Corr}}(n)&=& \sum_{n'} \,\rho(n') \, V_{\textrm{Corr}}(|n-n'|) 
    \quad \textrm{with}\\
  V_{\textrm{Corr}}(|n\!-n'\!|) &=&
    \sum_{j'} V_{ee}(nj,n'j')\! -\!
      \int  \frac{d^2\!R_{n'}\;e^2}{b^2\,\varepsilon
              \left|\vec R_{n}\!-\! \vec R_{n'} \right|}\; ,
              \nonumber
\end{eqnarray} 
where $b^2$ is the surface of the
two-dimensional unit cell and $\vec R_{n}$ lies in the plane
$n$. In
equation~(\ref{eq:UCorr}) we expect the largest corrections for $n'=n$ since
 the difference of the potential created by a uniformly and a
discretely charged plane is only sizeable at short distance. 
$V_{\textrm{Corr}}$ cannot be expressed analytically since it
involves an infinite sum over a two-dimensional lattice. The result from
numerical summation is given in Table~\ref{tab:VCorr}. The function
$V_{\textrm{Corr}}(\Delta n=|n-n'|)$ is rapidly decreasing and
we will only retain the $\Delta n=0$ term in the following. 
The correction term
then becomes
\begin{equation} \label{eq:UCorr2}
  U_{\textrm{Corr}}(n)=
       -\rho(n)\left(3.9\frac{e^2}{\varepsilon b}-U_0\right). 
\end{equation}
In fact, the second part of this correction overestimates the on-site interaction
$U_0$. The contribution of this term  is proportional to
$\rho(n)$ which reflects the fact that an electron feels its own
mean-field. This is an artifact of the mean-field approximation. In order to
improve our model, we require that the on-site interaction should only be
effective when there is more than one electron on the molecule
($\rho(n)>1$), leading to the correction term
\begin{equation} \label{eq:UCorr3}
  U_{\textrm{Corr}}(n)=
       -3.9\frac{e^2}{\varepsilon b}\rho(n)+U_0\max[0,\rho(n)-1]. 
\end{equation} 
\begin{table}
\begin{center}
\begin{tabular}{c | r @{$\,=\,$} l}
 $\Delta n=|n-n'|$ & 
    \multicolumn{2}{c}{$V_{\textrm{Corr}}(\Delta n)\;$(eV)} \\ \hline
     0             & $U_0-3.90\,e^2/\varepsilon b$&$
                      U_0-1.28=-0.28$ \\
     1             & $-0.042\,e^2/\varepsilon b$&$-0.014$\\
     2         & $-10^{-3}\,e^2/\varepsilon b$&$
                                    -3\times 10^{-4}$ 
\end{tabular}
\caption{
The function $V_{\textrm{Corr}}(\Delta n)$ as calculated 
by numerical summation.  }\label{tab:VCorr}
\end{center}
\end{table}
%

\begin{figure}
\begin{center}
\includegraphics[width=0.3\textwidth]{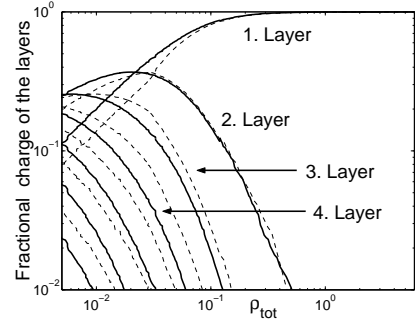}
\caption{
  Relative charging of the first layers as a function of total
  charge in the system (per
  area of the 2D unit cell). Both axis are in logarithmic units. 
  \emph{Solid line:} Conduction band.
  \emph{Dashed line:} Valence band. 
}\label{fig:ChargeProfile}
\end{center}
\end{figure}
\begin{figure}
\begin{center}
\hspace{0.5cm}
\begin{minipage}[c]{0.3\textwidth}
  \includegraphics[width=\textwidth]{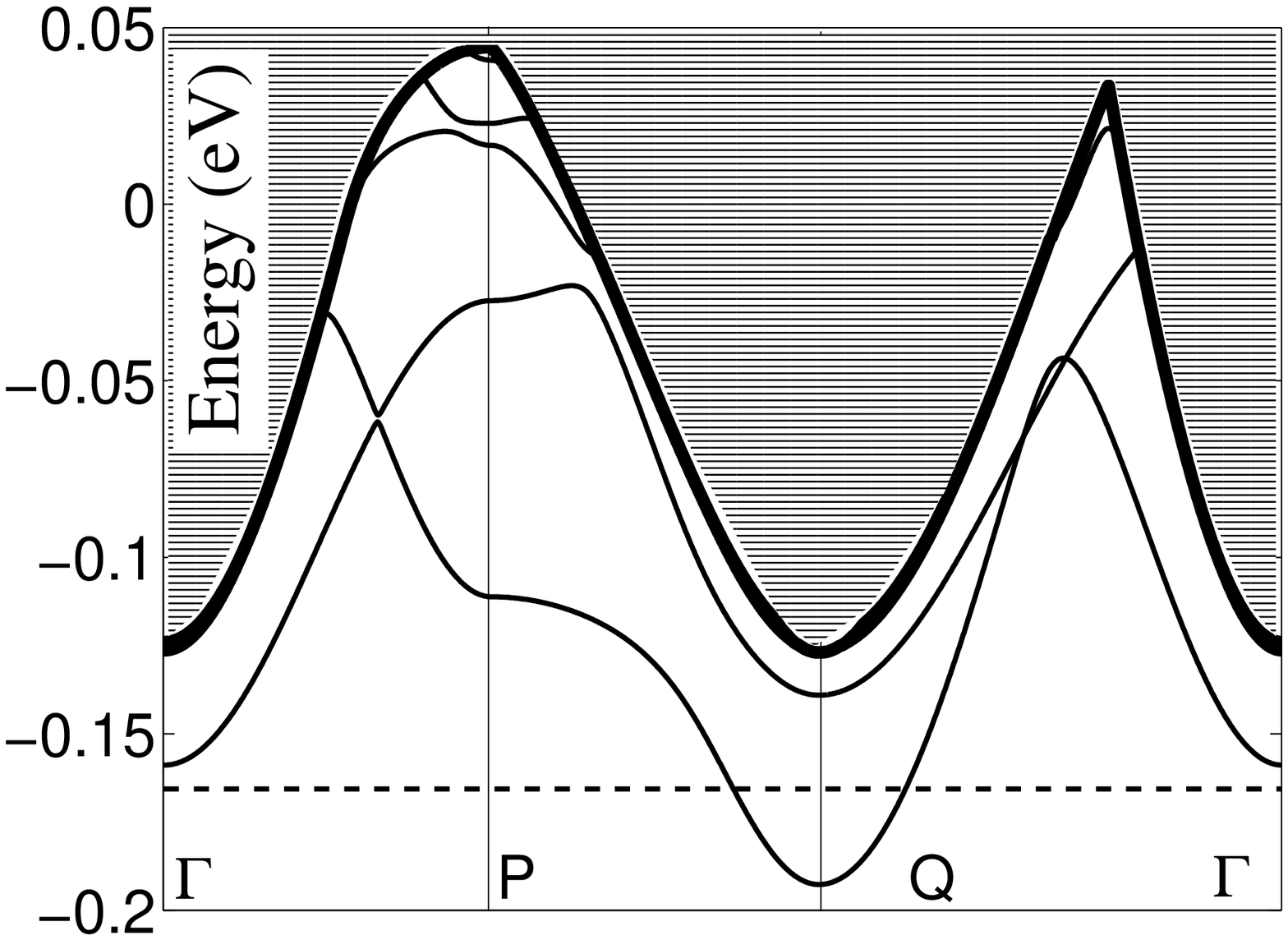}
\end{minipage}\vspace{0.2cm}\\
\caption{
 \emph{Solid lines:} Conduction bands below the continuous spectrum
 for $\rho_{\textrm{tot}}=0.1$. \emph{Bold solid line:} Onset of
  the continuous spectrum. \emph{Dashed line:} Fermi energy.
 }\label{fig:M2Dband}
\end{center}
\end{figure}
\begin{figure}
\begin{center}
(a)\hspace{0.5cm}
\begin{minipage}[c]{0.3\textwidth}
\includegraphics[width=\textwidth]{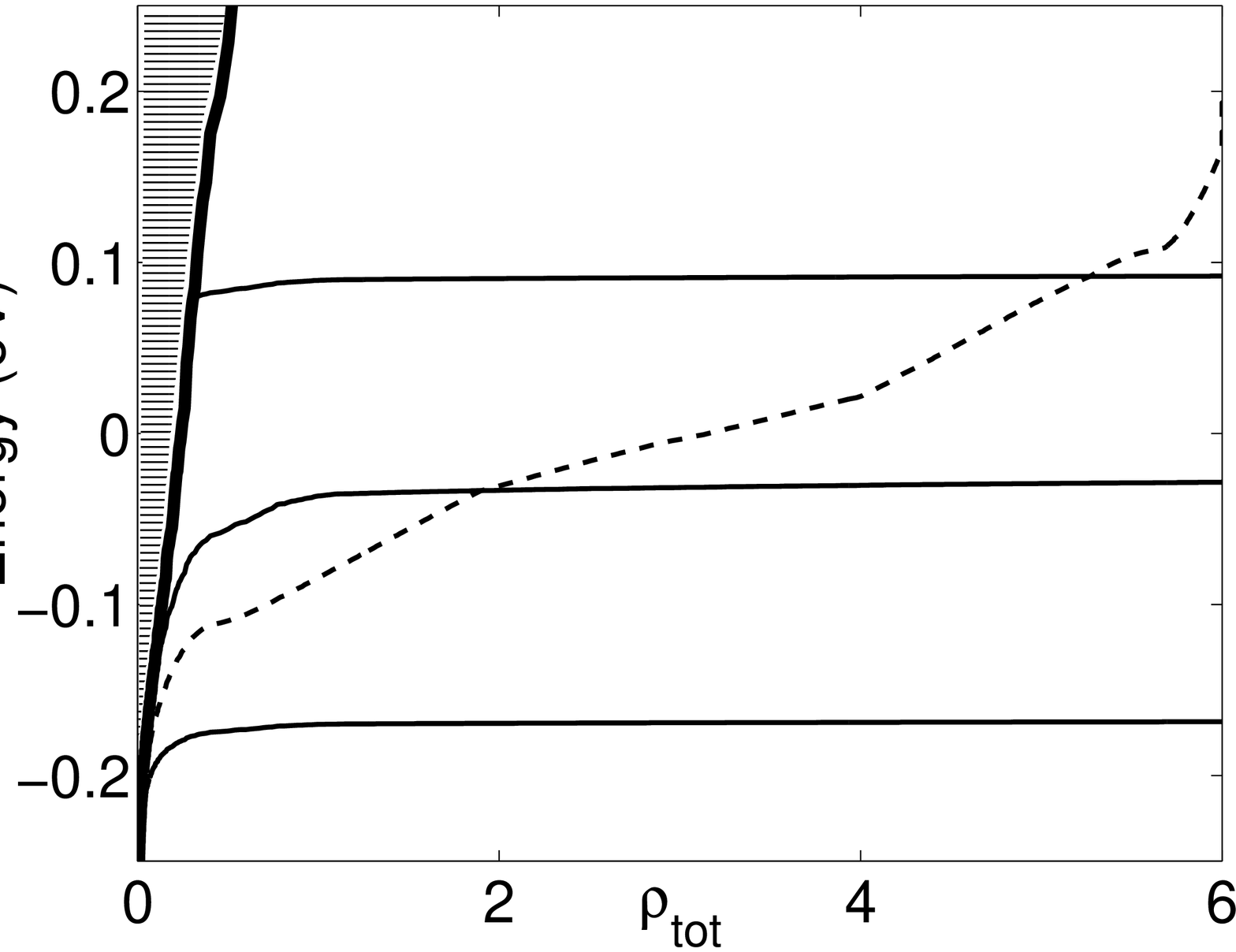}
\end{minipage}\\
(b)\hspace{0.5cm}
\begin{minipage}[c]{0.3\textwidth}
\includegraphics[width=\textwidth]{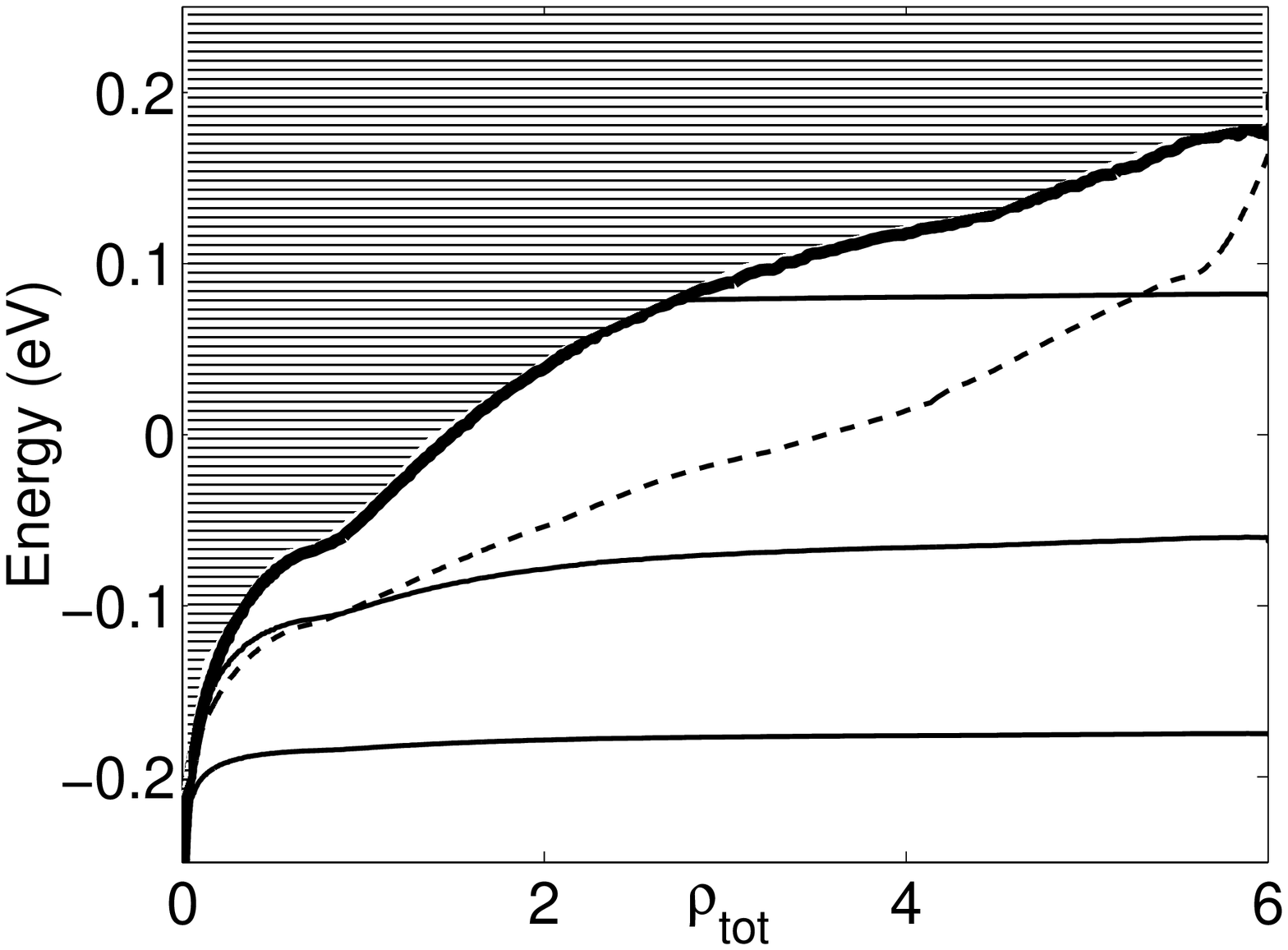}
\end{minipage}\vspace{0.2cm}\\
\caption{
(a) \emph{Solid lines:}  
 Discrete levels at Q as a function of $\rho_{\textrm{tot}}$. 
 \emph{Bold solid line:} Bottom of the continuous spectrum.
 \emph{Dashed line:} Fermi energy. 
(b) As (a), but without correction term $U_{\textrm{Corr}}$.
 }\label{fig:QEvol}
\end{center}
\end{figure}

The mean-field Hamiltonian was solved self-consistently at zero temperature
by numerical means. The mean-field potential~(\ref{eq:UMF2}) was used with the
correction term~(\ref{eq:UCorr3}). 
The resulting charge profile as a function of total
charge is shown in Fig.~\ref{fig:ChargeProfile}. We have done
the calculation for both, conduction and valence band, which led to similar
results. In the following we focus on the conduction band.
It can be seen that more than 98\% of the total charge is confined to the first
layer for doping higher than $\rho_{\textrm{tot}}=0.3$. 
Furthermore the
confinement to the interface increases with the total charge.           
The mean distance of the
charge distribution from the interface is defined as
$z_0 = \sum_{n \ge 1}\,n\,d\,\rho(n)/\rho_{\textrm{tot}}$, 
where $d=7$\AA{} is the
width of one layer. 
For $\rho_{\textrm{tot}}<0.1$,  this mean distance is
found to follow
a power law given by $z_0[\textrm{\AA}]\approx 3.8\,\rho_{\textrm{tot}}^{-1/3}$
with $\rho_{\textrm{tot}}$ in units of particles per area of the 2D unit cell.
This power law behavior with respect to $\rho_{\textrm{tot}}$ is identical to
the case of a standard space-charge layer
in a
continuous medium~\cite{art:ando82}, which predicts 
$z_0[\textrm{\AA}] = 3.02 \; (\rho_{\textrm{tot}}\,m_z/\varepsilon)^{-1/3}
=5.08\;\rho_{\textrm{tot}}^{-1/3}$, where
the effective mass (perpendicular to the interface) of
the bulk conduction band minima (X(1,0,0) points) is estimated to be
$m_z=0.92\,m_e$~\cite{art:satpathy92,art:laouini95}.
However, it is important to notice that our microscopic calculation gives a
different prefactor. In addition, $z_0$ tends to saturate for 
$\rho_{\textrm{tot}}>0.1$, a regime that cannot be understood in the
continuum model.
In order to test the role of the correction term, we repeated the same
calculation with $U_{\textrm{MF}}= U_{\textrm{Cap}}$, i.e. for uniformly
charged planes. This led to a somewhat weaker confinement at high doping
($\sim 10$\% less charge on the first layer for $\rho_{\textrm{tot}}=1$). At
low doping the effect of $U_{\textrm{Corr}}$ vanishes because the charge is
distributed over several layers.
Knowing the charge profile and hence $U_{\textrm{MF}}(n)$, the band structure can
be calculated. Fig.~\ref{fig:M2Dband} shows the conduction band for 
$\rho_{\textrm{tot}}=0.1$. Few 
bands lie below the continuous spectrum and only the lowest of the 3 LUMO-bands is
occupied in the region of the Q-point. In Fig.~\ref{fig:QEvol}(a)
 the evolution of the
discrete states at Q is shown as a function of doping. One observes that
the Fermi level is always below the bottom of the continuous spectrum. This is
true in general and reflects the fact that all electrons are bound to the
interface since the overall system is neutral. Furthermore, only the lowest
band is occupied at low doping which is again
consistent with theory of space-charge layers~\cite{art:ando82}. Of course, at
doping higher than $\rho_{\textrm{tot}}\approx 2$ the second LUMO-band has to be
filled as well. In addition, we note that the continuous spectrum shifts up
sharply as the total charge increases. Therefore, at doping higher than
$\rho_{\textrm{tot}}\approx 0.3$,  the first layer becomes 
essentially decoupled from the continuous states of the subsequent layers. 
The energy levels at Q remain constant and the dispersion is given by the
LUMO-bands of a single, isolated layer
(Fig.~\ref{fig:CondBand}(a)).
The decoupling effect is due to the
correction term~(\ref{eq:UCorr3}) which drastically shifts the energy of the
first layer with respect to the second one. 
If this term is omitted, as shown in 
Fig.~\ref{fig:QEvol}(b), then the bottom of the continuous spectrum    
follows roughly the Fermi energy and the decoupling of the first layer is much
less effective.
%
 
In A$_3$C$_{60}$, a scaling of the superconducting $T_c$ with increasing
lattice constant and hence with the DOS is observed~\cite{art:gunnarsson97}. 
However, the sharp peaks of the 2D DOS (Fig.~\ref{fig:CondBand}(b)) are not
reflected in the behavior of $T_c$ as measured by Sch\"on et
al.~\cite{art:schoen0}. A possible explanation for this is the
morahedral disorder which is considered in the following for the case of a
single 2D layer.

\begin{figure}
\begin{center}
\includegraphics[width=0.3\textwidth]{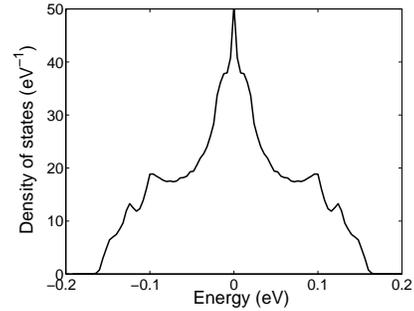}\vspace{0.2cm}
\caption{
DOS for a disordered 50x50 site lattice and a flipping ratio of
15\%. The statistical average
over 200 disorder configurations was taken. 
}\label{fig:DisDOS}
\end{center}
\end{figure}

The rotational motion of a C$_{60}$ molecule in a crystal is described by a
potential in the three 
Euler angle coordinates. This potential has, in addition to an absolute
minimum,  a local minimum which is only
$\Delta=11$~meV higher and corresponds to a non-equivalent
orientation of the molecule~\cite{art:david93}. 
Therefore, C$_{60}$ molecules  can flip into
the second orientation causing the so-called morahedral disorder. 
The ratio between the number of flipped and unflipped
molecules is given by a Boltzmann factor $\exp(-\Delta/kT)$.
However, a freezing in is observed
at 85~K. Below this temperature the crystal is in a glass phase with 15\% of
the molecules being flipped. 
If the problem is
considered from a tight-binding point of view, then one expects
the hopping integrals to be different for hopping between inequivalently
oriented molecules. Hence, one gets a
tight-binding model where different hopping integrals are
distributed randomly. This will suppress structure in the DOS, as was calculated
 by Gelfand and 
Lu in the case of bulk C$_{60}$~\cite{art:gelfand92b}.
Here, we consider the effect of morahedral disorder on the DOS
 of a single electron-doped 
[001] layer. The two types of inequivalent
 orientations lead to two sets of LUMO wave-functions. We shall denote the new
 ones on flipped molecules as $\ket{\tilde x}$, $\ket{\tilde y}$ and $\ket{\tilde z}$. These two sets can be related
 by
 a 90-degree rotation around the z-axes. 
The hopping integrals between
inequivalently oriented molecules are given in Table~\ref{tab:HopIntDis}.
In order to calculate the DOS of the disordered system we set  
up a finite, two dimensional
system which could be diagonalized exactly (for non-interacting electrons). The
resulting DOS is shown in Fig.~\ref{fig:DisDOS}. The van-Hove
singularities are washed out, but a rather sharp peak remains 
at the center of the  band. 
\begin{table}
\begin{center}
\begin{tabular}{c|ccc}
&  $\ket{x}$ & $\ket{y}$ &  $\ket{z}$ \\
\hline 
$\bra{\tilde y}$ & $\tilde t_{xy}=22.9$  & $\pm \tilde t_{yy}=40.8$  & 
$0$ \\
$\bra{\tilde x}$ &  $\pm \tilde t_{xx}=19.2$  & $\tilde t_{xy}=22.9$      
     & $0$ \\
$\bra{\tilde z}$ & $0$              & $0$           & $\tilde
t_{zz}=-29.4$ 
\end{tabular}
\caption{
Hopping integrals (in meV) for hopping from an
unflipped to a flipped molecule in the $(110)$ direction of a fcc
lattice. The new (tilded) wave-functions were ordered according to 
their parity.
Note that the sign of $\tilde t_{xx}$ and $\tilde t_{yy}$  change for hopping in
the $(1\bar{1}0)$ direction. Numerical values were taken
from~\cite{art:gelfand92b} and adjusted by a factor~11 as proposed
by Gelfand and Lu.   }\label{tab:HopIntDis}
\end{center}
\end{table}
%


In conclusion, we have investigated the electronic properties of surface doped C$_{60}$. The
charge profile was calculated as a function of total charge in the system and it
was found that the charge accumulation on the first layer increases 
with the total charge. At high doping, above $\sim 0.3$ electron per
molecule, the first layer becomes essentially decoupled from the subsequent layers.
This suggests that the electronic system should be well described by a
single layer in this regime. In particular, the relevant DOS is then the one
of a
two dimensional system and hence substantially higher than in the bulk. Having
chosen the unidirectional structure (one orientation per primitive cell), we found
 maximal values of the DOS near the center of the band. 
Introducing morahedral disorder led to a smoothened DOS which still
shows a peak at
half filling, i.e. in the region where Sch\"on et al. measure the maximum
$T_c$. 

\begin{acknowledgement}
  We thank B.\ Batlogg for fruitful discussions throughout this
  work. D.\ P.\ also acknowledges support from the Center for Theoretical
  Studies and from the Physics Department at ETH-Z\"urich. 
\end{acknowledgement}

\end{document}